\documentclass[aip,amsmath,amssymb,reprint]{revtex4-1}

\usepackage{graphicx}
\usepackage{dcolumn}
\usepackage{bm}
\usepackage[utf8]{inputenc}
\usepackage[T1]{fontenc}
\usepackage{mathptmx}
\usepackage{etoolbox}
\usepackage{enumitem}
\usepackage{amsmath}
\usepackage[table]{xcolor}
\usepackage{booktabs} 
\usepackage{booktabs}
\usepackage{multirow}
\usepackage{makecell}
\usepackage{siunitx}
\usepackage{hyperref}
\usepackage{booktabs}
\usepackage{xcolor}
\usepackage{tikz}
\usetikzlibrary{positioning, calc, arrows.meta}
\usepackage{bm} 
\PassOptionsToPackage{normalem}{ulem}
\usepackage{ulem}

\providecolor{added}{rgb}{0,0,1}
\providecolor{deleted}{rgb}{1,0,0}

\makeatletter
\def\@email#1#2{%
 \endgroup
 \patchcmd{\titleblock@produce}
  {\frontmatter@RRAPformat}
  {\frontmatter@RRAPformat{\produce@RRAP{*#1\href{mailto:#2}{#2}}}\frontmatter@RRAPformat}
  {}{}
}%
\makeatother

\newcommand{\fwd}{f}           

\begin{document}

\preprint{AIP/123-QED}

\title[XXXXX]{Neural Network Based Molecular Structure Retrieval from Coulomb Explosion Imaging Data}

\author{A.~Ghanaatian}
\affiliation{Computer Science Department, Kansas State University, Manhattan, KS 66506, USA}
\author{A.~K.~Ravi}
\affiliation{Department of Mathematics, Kansas State University, Manhattan, KS 66506, USA}
\author{J.~Stallbaumer}
\affiliation{James R. Macdonald Laboratory, Department of Physics, Kansas State University, Manhattan, KS 66506, USA}
\author{H.~V.~S.~Lam}
\affiliation{James R. Macdonald Laboratory, Department of Physics, Kansas State University, Manhattan, KS 66506, USA}
\author{A.~Rudenko}
\affiliation{James R. Macdonald Laboratory, Department of Physics, Kansas State University, Manhattan, KS 66506, USA}
\author{L.~Greenman}
\affiliation{James R. Macdonald Laboratory, Department of Physics, Kansas State University, Manhattan, KS 66506, USA}
\author{N.~Albin}
\affiliation{Department of Mathematics, Kansas State University, Manhattan, KS 66506, USA}
\author{D.~Caragea}
\affiliation{Computer Science Department, Kansas State University, Manhattan, KS 66506, USA}
\author{D.~Rolles}
\affiliation{James R. Macdonald Laboratory, Department of Physics, Kansas State University, Manhattan, KS 66506, USA}
\email{rolles@ksu.edu}

\date{\today}

\begin{abstract} 
Determining the structure and following the structural evolution of molecules undergoing chemical reactions is one of the key goals of ultrafast molecular physics and chemistry. Recently, Coulomb explosion imaging has emerged as a promising technique for imaging the evolving structure of individual molecules in the gas phase. However, its practical application for structure determination is hampered by the lack of suitable algorithms to directly retrieve the molecular structure from the measured fragment-ion momentum data.
Here, we propose a scheme to solve the underlying inverse problem by employing neural networks to infer the initial atomic positions from the final ion momenta on an event-by-event basis. 
Using this scheme, we retrieve the structure of several polyhalomethane isomers from simulated Coulomb explosion imaging data with an average per-atom position error of approximately 0.1 atomic units, i.e.\ to within 5\% of the typical bond lengths. This development paves the way for an automated structure retrieval from Coulomb explosion data one molecule at a time, making it ideally suitable for analyzing pump-probe experiments where several products are formed that need to be distinguished.
\end{abstract}

\maketitle

\begin{quotation}
XXXXXX
\end{quotation}


\section{Introduction}
Ultrafast dynamics in electronically excited states of organic molecules drive important photochemical processes such as the photosynthesis in plants \cite{Cheng2009}, human vision \cite{Polli2010} or photosynthesis of vitamin D \cite{Havinga1961}, as well as a multitude of important atmospheric reactions \cite{Papanastasiou2014, SaizLopez2012}. Despite tremendous progress in the fields of ultrafast experimental methods and \textit{ab initio} simulations, understanding ultrafast photochemical reactions still poses significant challenges. They often involve non-adiabatic dynamics, i.e., correlated motion of the electrons and nuclei outside the Born-Oppenheimer approximation. A successful approach for improving our understanding of ultrafast photochemistry is to investigate isolated molecules in the gas phase since this facilitates the direct comparison of experimental measurements with high-level molecular dynamics simulations of ultrafast photochemical processes. 

Ultrafast photochemical reactions in the gas phase have been investigated by a large number of pump-probe spectroscopic methods. Many of the ultrafast spectroscopies are significantly more sensitive to changes in the electronic structure of a molecule than to its structural dynamics. However, the correlated nature of electronic and nuclear motion during non-adiabatic dynamics requires an equally detailed investigation of the structural evolution. Novel femtosecond time-resolved imaging techniques enabled, e.g., by ultrashort X-ray free electron laser (XFEL) \cite{stankus2020advances, odate2023brighter} and electron sources \cite{centurion2022ultrafast} have demonstrated the substantial benefit of direct measurement of the nuclear geometry evolution during non-adiabatic processes. Recently, Coulomb explosion imaging (CEI), originally realized through collisions of fast ion beams with thin foils\cite{vager1989coulomb, herwig2013},  has emerged as another popular method for investigations of ultrafast photochemical dynamics in individual molecules, especially in cases that involve multiple channels with diverse structural changes \cite{Li2025} or product structures with significant hydrogen rearrangement \cite{Green2025}, which are difficult to distinguish by X-ray and electron diffraction techniques. In Coulomb explosion imaging, a gas-phase molecule is highly ionized by an ultrashort intense laser pulse - generated, e.g., by an X-ray free-electron laser (XFEL)\cite{li2021simple,boll2022x,li2022coulomb,Li2025,Green2025} or a near-infrared femtosecond laser\cite{pitzer2013direct,Kling2019, endo2020capturing,bhattacharyya2022strong, lam2024differentiating, venkatachalam2025ch2i2}. Coulomb repulsion rapidly “explodes” the molecular ion, and direct information about the molecular structure before the explosion is encoded in the momentum distribution of the ion fragments based on the inverse relation between the internuclear distance and Coulomb repulsion. 

A combination of the CEI technique with coincidence detection of the fragment ions with full three-dimensional momentum resolution was shown to be applicable to a wide range of molecules \cite{ablikim2016identification, Pathak2020, endo2020capturing, bhattacharyya2022strong, boll2022x, li2022coulomb, Severt2024}, including, in particular, the capability to identify and distinguish molecular isomers using CEI induced by strong-field ionization \cite{pitzer2013direct, Kling2019, endo2020capturing, McDonnell2020, bhattacharyya2022strong, lam2024differentiating, Lam_SO2, venkatachalam2025ch2i2} or soft X-ray inner-shell ionization \cite{ablikim2016identification, Ablikim2017, Pathak2020}. The latter improved tremendously when utilizing intense, few-femtosecond X-ray pulses from a free-electron laser, which has provided snapshots of the molecular structure of polyatomic molecules with unprecedented clarity \cite{boll2022x, li2022coulomb, Green2025}. 

However, while straightforward for diatomic molecules, the retrieval of the molecular structure from the fragment-ion momentum distribution becomes extremely challenging as the size of the molecule increases. Frequently, the molecular structure at the time of ionization is inferred by comparing the experimentally measured fragment-ion momentum correlations with simulations of the Coulomb explosion process for various possible molecular geometries\cite{Li2025,Green2025,endo2020capturing, bhattacharyya2022strong, lam2024differentiating, venkatachalam2025ch2i2}. In that way, CEI often provides a set of detailed constraints on the possible geometries instead of directly yielding the structure of the investigated molecules. Moreover, as the size of the molecule increases, the data become extremely high dimensional, and obtaining quantitative information on the molecular geometry by manually comparing many particular subsets of the data to the corresponding simulated predictions becomes more and more challenging. Therefore, more powerful and general methods for structure retrieval from CEI data are needed. Machine learning approaches, including traditional neural networks and novel powerful deep neural networks, have been increasingly applied to various problems in physical chemical in recent years \cite{Westermayr2021,Dral021}, and seem well suited candidates for this task. For example, it was shown that by applying dimensionality reduction and clustering methods to CEI data of a mixed sample of six-atom molecules, complete differentiation of molecular structures in the sample is possible, even if the structures are not directly reconstructed \cite{venkatachalam2025exploiting}. It was also demonstrated that molecular structure reconstruction from CEI data can be achieved using deep neural networks trained on simulated Coulomb explosion data under the assumption that only one molecular species is present in the data \cite{Li2025arxiv,guo2025decoding}. Here we address the challenge of direct molecular geometry reconstruction from CEI data for multiple structures that may be present in the data of a time-resolved photochemistry experiment. We show that by combining CEI with a deep neural network-based approach, we can retrieve the molecular structure of several five-atom molecules in a mixed sample solely from their simulated fragment-ion momentum distributions, even identifying an `unexpected' molecular structure that was not included in the network's training data. As a practical example to test our method, we chose the chiral molecule CHBrClF and several of its (fictitious) isomers because it consists of five unique atoms and was the target of a prominent 5-body CEI experiment reported in the literature \cite{pitzer2013direct}.

\begin{figure}
    \centering
    \includegraphics[width=\columnwidth]{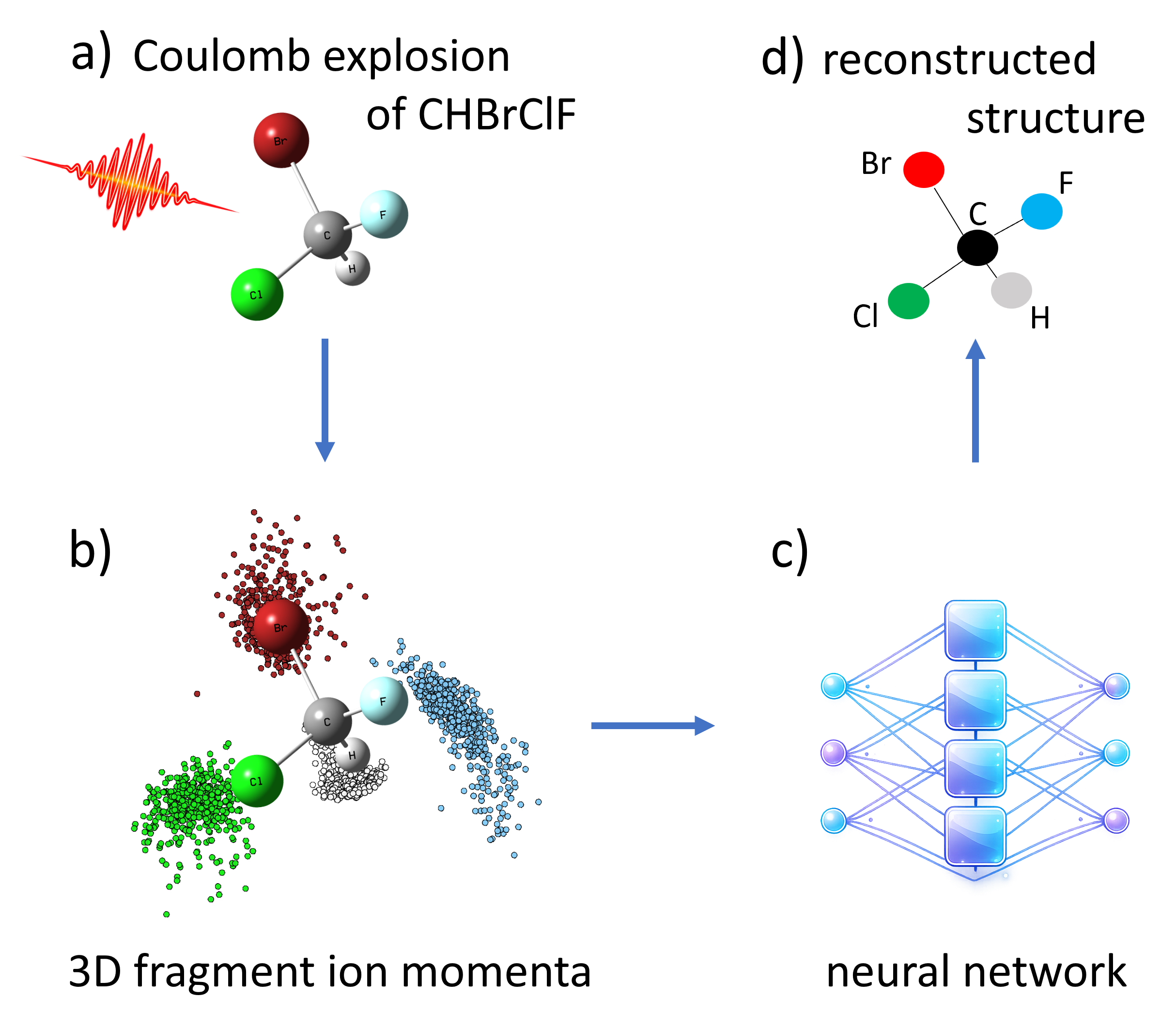}
    \caption{Schematic of the molecular structure reconstruction approach: A neural network is used to reconstruct the molecular structure of various CHBrClF isomers from the three-dimensional fragment-ion momenta resulting from the Coulomb explosion induced by a femtosecond laser pulse.}
    \label{fig:schematic}
\end{figure}


\section{Neural Networks for Solving Inverse Problems}
\label{sec:related}

Mathematically, retrieving the molecular geometry from CEI data represents a typical example of an inverse problem. Inverse problems have a long history in computational physics and applied mathematics. Classical regularization methods, particularly Tikhonov regularization~\citep{tikhonov1977}, stabilize ill-posed inverse maps by adding penalty terms that enforce smoothness or other desirable properties on solutions. Variational methods frame inversion as optimization problems with explicit regularization functionals~\citep{vogel2002}. These deterministic approaches provide computationally efficient point estimates, but they offer limited uncertainty quantification and tend to struggle when the solution space is multi-modal.

As an alternative approach, Bayesian formulations~\citep{kaipio2006, stuart2010} cast inverse problems as posterior inference over unknown parameters given the observed data. This explicitly represents uncertainty and allows prior knowledge to be incorporated, yet practical inference can become challenging in high dimensions, especially when posteriors are multi-modal or the forward model is expensive to evaluate. Simulation-based inference methods have emerged to address some of these computational bottlenecks~\citep{vogel2002}.

Building on this, modern generative neural networks offer flexible representations that scale to high-dimensional settings. Variational autoencoders (VAEs)~\citep{kingma2014} learn the latent structure of complex data by encoding it into a continuous, probabilistic latent space, enabling the generation of new, unique samples that are similar to the training data. Normalizing flows~\citep{dinh2017} transform a simple probability distribution  into a complex target distribution by applying a series of invertible and differentiable functions, a process that allows for the exact  likelihood evaluation. These generative models excel at representing multi-modal posteriors and can scale to high-dimensional problems where traditional methods struggle.

Complementing these approaches, physics-informed learning embeds known constraints into training objectives or architectures~\citep{raissi2019}, while equivariant designs encode symmetries directly into the network structure to improve sample efficiency and generalization~\citep{satorras2021}. Together, these developments bridge data-driven and physics-based perspectives, offering improved stability and interpretability for scientific inverse problems.

Building on prior methodological advances and considering the ill-posed and multi-modal nature of molecular structure retrieval from CEI data, we adopt two complementary neural network formulations. A deterministic multilayer perceptron (MLP) is used to learn a direct inverse mapping from fragment-ion momenta to initial atomic positions, providing an efficient point-estimate solution analogous to classical regularized inversion. To explicitly account for one-to-many mappings in which similar momentum configurations arise from distinct molecular geometries, we additionally employ a variational autoencoder (VAE) that learns to map input data to a continuous, probabilistic latent space. This combined approach enables a direct comparison between deterministic and probabilistic inversion strategies under identical physical constraints.

\section{Methods}
\label{sec:methods}

Since the large data sets needed for training neural networks are currently still difficult to generate purely experimentally, and since Coulomb explosion simulations allow for more flexibility in the molecular geometries used for the training data, we have trained and tested the networks in the present work solely on simulated data generated as described in Section \ref{CEsimulation}. Although there are well-known quantitative discrepancies between the fragment ion momenta obtained from classical Coulomb explosion simulations and those measured experimentally \cite{bhattacharyya2022strong,boll2022x,li2022coulomb, venkatachalam2025exploiting}, these differences do not affect the fundamental conclusions of the present work nor the comparison of the performance of the MLP and VAE networks, described in Section \ref{sec:models}. Some possible schemes to generalize networks trained on simulated data to be applicable to actual experimental data are discussed in Refs.~\cite{Li2025arxiv,guo2025decoding}. As powerful new X-ray and near-infrared laser sources with repetition rates of 100 kHz or more are becoming more prevalent, the experimental efforts and time investment needed to generate large experimental data sets for network training purposes will decrease, making it feasible in the near future to also use experimental data for training. 

\subsection{Numerical Simulation of the Coulomb Explosion Process}
\label{CEsimulation}

To generate the Coulomb explosion data for training and testing the networks, we apply a simple numerical approach simulating an instantaneous explosion process as classical motion of point charges in a purely Coulombic potential, which is commonly used in CEI studies 
\cite{bhattacharyya2022strong,boll2022x,li2022coulomb,endo2020capturing,lam2024differentiating,ablikim2016identification, Ablikim2017, Pathak2020,Lam_SO2,venkatachalam2025ch2i2,Li2025,Green2025,venkatachalam2025exploiting,Li2025arxiv,guo2025decoding}.

Upon acquiring a charge $q_i$ at each atomic site $i$ of the molecule, the motion of the resulting ionic fragments with mass $m_i$ and position $\mathbf{r}_i$ can be described approximately (assuming instantaneous ionization to the final charge state, instantaneous localization of the charges, and a purely Coulombic potential energy surface) by a set of coupled classical equations of motion based on the electrostatic Coulomb force for a system of point charges:
\begin{equation}
\label{eq:forwardpath}
m_i  \frac{d^2 \mathbf{r}_i}{dt^2}= \sum_{j\neq i}  \frac{q_i q_j}{|\mathbf{r}_i- \mathbf{r}_j |^3}(\mathbf{r}_i -\mathbf{r}_j)	
\end{equation}

Under these assumptions, the measured final fragment momenta $\mathbf{p}_{i,f}=\displaystyle \lim_{t\to\infty}\mathbf{p}_i$ are uniquely determined by the initial atomic positions at time $t=0$, $\mathbf{r}_{i,0}$, thus providing a momentum-space “image” of the initial molecular geometry.

\textbf{ML Task.} From the machine learning viewpoint, solving Eq.~\eqref{eq:forwardpath} provides a function
\begin{equation}
\label{eq:forwardmodel}
\mathbf{y} = \fwd(\mathbf{x}),
\end{equation}
where $\mathbf{x} = (\mathbf{r}_{1,0},\ldots,\mathbf{r}_{N_{\text{atoms}},0})$ is the tall $3N_{\text{atoms}}$-vector formed by stacking the initial positions of all atoms together, and $\mathbf{y}=(\mathbf{p}_{1,f},\ldots,\mathbf{p}_{N_{\text{atoms}},f})$ is the corresponding vector formed by stacking the final momenta. Our goal in this paper is to train a neural network model to invert this function--mapping the final momentum measurements, $\mathbf{y}$ back to the corresponding initial configuration $\mathbf{x}$.

\label{sec:data}
\textbf{Data Generation.}  
We synthesize paired samples \((\mathbf{y},\mathbf{x})\) by drawing initial atomic positions \(\mathbf{x}\) from artificially generated ensembles of molecular geometries and calculating the corresponding final fragment momenta through simulation as described above. The ensembles of molecular geometries are detailed in Section \ref{sec:results} and include atoms placed randomly within a sphere of radius \(R\) as well as several idealized isomers of a five-atom molecule.
The fragment momenta are obtained deterministically by solving Eq.~\eqref{eq:forwardpath} numerically.
For the results reported, we used the RK45 method in SciPy~\cite{2020SciPy-NMeth} to solve the ordinary differential equation system in the time interval $[0, 10^{10}]$ a.u.\ with the default tolerances ($10^{-3}$ for relative tolerance and $10^{-6}$ for absolute tolerances). 

\textbf{Coordinate rotation and reference fixing.}
For an $N$-atom molecule, both $\mathbf{x}$ and $\mathbf{y}$ consist of $3N$ scalar quantities. From the perspective of machine learning, however, this is not an efficient representation of the data. It ignores the rotational equivariance and translation invariance of the forward operator $f$, meaning that data points for very similar molecular structures may be arbitrarily far apart simply because they are rotated and translated differently.

To mitigate this problem, we apply a well-defined transformation to the data, eliminating three scalar quantities from both input and output. In the case of CHBrClF (Section~\ref{sec:results}), we choose the following change of variables through a combination of translation and rotation for all positions and momenta: 
\begin{itemize}
\item The initial position of carbon, $\tilde{\mathbf{r}}_{C,0}$ is the origin.
\item The final momentum of carbon, $\tilde{\mathbf{p}}_{C,f}$ points in the positive $x$-direction.
\item The final momentum of bromine, $\tilde{\mathbf{p}}_{Br,f}$ lies in the $xy$-plane and has a non-negative $y$ coordinate.
\end{itemize}
The same rotation as for the momenta is also applied to the initial positions, and the three zeroed quantities in both $\tilde{\mathbf{x}}$ and $\tilde{\mathbf{y}}$ are then removed.

While certain machine learning approaches known as Geometry-Informed Neural Networks (GINN) are specifically designed to respect the symmetries of physical systems~\cite{brandstetter2022message,thomas2018tfn}, the neural network architectures we have explored perform significantly better when such rotational and translation invariances are removed prior to training and testing. Since the procedure described above is straightforward to implement, we therefore chose to perform it for all models to allow for a fair comparison of the performance across different models.
Note that this reduction in dimension is not minimal--for example, additional reductions can be made by using the conservation of momentum and transforming to more suitable internal (e.g.\ Jacobi) coordinates \cite{Severt2024}. However, for the sake of maintaining a more intuitive representation for our case study, we opted to stay in Cartesian coordinates.

\subsection{Application of Neural Networks for Molecular Structure Retrieval}\label{sec:models}

We have applied two neural network architectures for the task of inverting the function $f$ in Eq.~\eqref{eq:forwardmodel}. The goal is to recover initial position data, $\mathbf{x}$, from final momentum data, $\mathbf{y}$. This section describes the specific models used in our experiments. To fix notation and establish the terminology used throughout this paper, we begin by describing the architecture of a simple multilayer perceptron (MLP), which serves as the foundational model for many of the neural networks that are commonly used in practical applications.

\subsubsection{Multilayer perceptron (MLP)}
\label{MLP}

\begin{figure}[b]
    \centering
\tikzstyle{mlp-block}=[fill=blue!10]
\tikzstyle{layer-conn}=[line width=2pt,-stealth, color=black!30]
\tikzstyle{param-inp}=[line width=1pt,-stealth, color=black!30, dashed]
\begin{tikzpicture}[scale=0.85, transform shape]

\begin{scope}[shift={(-2.5,0)}]
\fill[mlp-block] (0,1) rectangle (1,5);
\node at (0.5,5.5) {$\mathbf{h}^{(0)}$};
\node at (0.5,4.5) {$h^{(0)}_1$};
\node at (0.5,3.75) {$h^{(0)}_2$};
\node at (0.5,3) {$\vdots$};
\node at (0.5,1.5) {$h^{(0)}_{n_0}$};
\end{scope}

\begin{scope}[shift={(-0.5,0)}]
\fill[mlp-block] (0,0) rectangle (1,6);
\node at (0.5,6.5) {$\mathbf{h}^{(1)}$};
\node at (0.5,5.5) {$h^{(1)}_1$};
\node at (0.5,4.75) {$h^{(1)}_2$};
\node at (0.5,3) {$\vdots$};
\node at (0.5,0.5) {$h^{(1)}_{n_1}$};
\end{scope}

\node (v1) at (-3.5,3) {};
\node (v2) at (-2.5,3) {};
\draw[layer-conn]  (v1) edge (v2);

\node (v1) at (-1.5,3) {};
\node (v2) at (-0.5,3) {};
\draw[layer-conn]  (v1) edge (v2);

\node (v1) at (0.5,3) {};
\node (v2) at (1.5,3) {};
\draw[layer-conn]  (v1) edge (v2);

\node (v1) at (2.5,3) {};
\node (v2) at (3.5,3) {};
\draw[layer-conn]  (v1) edge (v2);

\node (v1) at (4.5,3) {};
\node (v2) at (5.5,3) {};
\draw[layer-conn]  (v1) edge (v2);

\begin{scope}[shift={(1.5,0)}]
\fill[mlp-block] (0,0) rectangle (1,6);
\node at (0.5,6.5) {$\mathbf{h}^{(2)}$};
\node at (0.5,5.5) {$h^{(2)}_1$};
\node at (0.5,4.75) {$h^{(2)}_2$};
\node at (0.5,3) {$\vdots$};
\node at (0.5,0.5) {$h^{(2)}_{n_{2}}$};
\end{scope}

\begin{scope}[shift={(3.5,0)}]
\fill[mlp-block] (0,1) rectangle (1,5);
\node at (0.5,5.5) {$\mathbf{h}^{(3)}$};
\node at (0.5,4.5) {$h^{(3)}_1$};
\node at (0.5,3.75) {$h^{(3)}_2$};
\node at (0.5,3) {$\vdots$};
\node at (0.5,1.5) {$h^{(3)}_{n_3}$};
\end{scope}

\node at (-3.5,3) {$\mathbf{y}$};

\node (v3) at (-1,4) {$W^{(1)}$};
\node (v11) at (-1,2) {$\mathbf{b}^{(1)}$};

\node (v5) at (1,4) {$W^{(2)}$};
\node (v12) at (1,2) {$\mathbf{b}^{(2)}$};

\node (v9) at (3,4) {$W^{(3)}$};
\node (v14) at (3,2) {$\mathbf{b}^{(3)}$};

\node at (5.5,3) {$\mathbf{\hat{x}}$};

\node (v4) at (-1,3) {};
\node (v6) at (1,3) {};
\node (v10) at (3,3) {};
\draw[param-inp]  (v3) edge (v4);
\draw[param-inp]  (v5) edge (v6);
\draw[param-inp]  (v9) edge (v10);
\draw[param-inp]  (v11) edge (v4);
\draw[param-inp]  (v12) edge (v6);
\draw[param-inp]  (v14) edge (v10);
\end{tikzpicture}
\caption{\label{fig:mlp} Schematic representation of a fully connected Multilayer Perceptron (MLP) architecture. The network consists of an input layer that accepts the input data $\mathbf{y}$. The data are transformed through a sequence of layers via Eq.~\eqref{eq:layer-transformation} to produce the final output $\mathbf{\hat{x}}$. The weights, $W^{(k)}$ and biases, $\mathbf{b}^{(k)}$ control how each layer transforms its inputs. These parameters are learned during training.}
\end{figure}
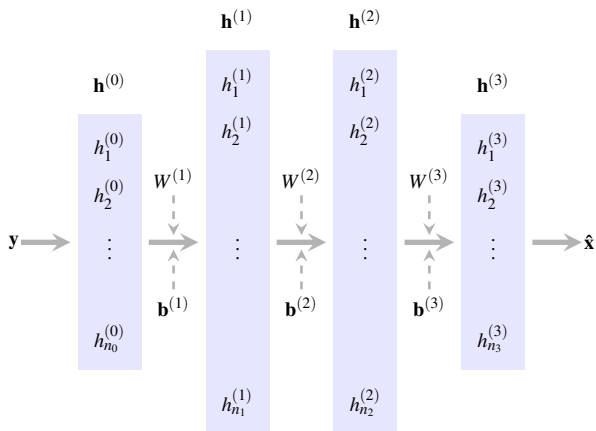

Multilayer perceptrons are feed-forward neural networks consisting of multiple layers of interconnected neurons that learn non-linear mappings from inputs to outputs through backpropagation. An MLP can be visualized as a collection of  neurons organized into layers (see Fig.~\ref{fig:mlp}). These layers include an input layer, an output layer, and one or more hidden layers. The feed-forward nature of an MLP refers to the fact that each neuron is directly influenced only by the outputs of the neurons of the preceding layer.

Given the vector, $\mathbf{y}$, of final momenta, an MLP with $L$ layers produces a prediction, $\hat{\mathbf{x}}$, of the initial positions by a composition of functions. Let $\mathbf{h}^{(0)}=\mathbf{y}$ be the input. The MLP performs a sequence of non-linear transformations,
\begin{equation}\label{eq:layer-transformation}
    \mathbf{h}^{(k)} = \sigma^{(k)}(W^{(k)}\mathbf{h}^{(k-1)} + \mathbf{b}^{(k)}),\quad\text{for }k=1,2,\ldots,L,
\end{equation}
to produce the final output $\hat{\mathbf{x}}=\mathbf{h}^{(L)}$. The number of layers, $L$, as well as the number of neurons in a  layer are hyperparameters that define the MLPs structure. If we let $n_k$ be the number of neurons in the $k$th layer, then the various quantities in Eq.~\eqref{eq:layer-transformation} have the following dimensions.
\begin{equation*}
    \mathbf{h}^{(k)},\mathbf{b}^{(k)}\in\mathbb{R}^{n_k},\quad
    W^{(k)}\in\mathbb{R}^{n_{k}\times n_{k-1}}.
\end{equation*}
Conceptually, all of the entries of the weight matrices $W^{(k)}$ and bias vectors $\mathbf{b}^{(k)}$ can be collected together into a set of model parameters, $\theta$, which are updated during training. The activation functions $\sigma^{(k)}$ are generally simple, nonlinear, elementwise functions, such as a rectified linear unit (ReLU) or sigmoid function.

Training the neural network involves iteratively updating the weights and biases through some variant of gradient descent on a cost function, $J(\theta)$, that evaluates the ability of the network with model parameters $\theta$ to predict the initial position data correctly. For the problem of molecular reconstruction, a natural candidate is the mean squared error loss. For a given momentum-position pair, $(\mathbf{y},\mathbf{x})$, in the training dataset and the corresponding MLP prediction, $\hat{\mathbf{x}}$, the MSE loss function is defined as
\begin{equation}
    \mathcal{L}_{\text{MSE}}(\mathbf{x},\hat{\mathbf{x}})
    = \|\mathbf{x}-\hat{\mathbf{x}}\|_2^2,
    \label{loss}
\end{equation}
the square of the Euclidean distance between the true and predicted positions. The average loss across all $m$ training data points,
\begin{equation}
    J(\theta) = \frac{1}{m}\sum_{i=1}^{m}
    \mathcal{L}_{\text{MSE}}(\mathbf{x}^{(i)},\hat{\mathbf{x}}^{(i)}),
        \label{loss-all}
\end{equation}
is called the cost function, and is thought of as a function of the model parameters. The MLP is trained by gradient descent (or one of its variants) on the cost function $J$.

While this description overlooks several technical points (e.g., mini-batching), it captures the broad idea shared by the neural network models in this section. Each model is a function with many parameters ($\theta$) that is trained to predict initial position data $\mathbf{x}$ from final momentum data $\mathbf{y}$   by means of an iterative optimization algorithm applied to some measure of loss $J(\theta)$ between the predictions $\hat{\mathbf{x}}$ and desired values $\mathbf{x}$.

\subsubsection{Variational Autoencoders (VAE)} \label{Generative}

Given the potential for degeneracies in mapping from momentum back to initial configurations, we assess the suitability of the MLP by comparing its performance to a Variational Autoencoder (VAE). A VAE consists of two interconnected neural networks: an encoder and a decoder. 

Rather than mapping the input $\mathbf{y}$ directly to a single output $\mathbf{x}$, the encoder maps it to a latent probability distribution. Specifically, it outputs the parameters of a multivariate normal distribution in a lower-dimensional latent space: a mean ($\mu$) and a variance ($\sigma^2$). It then draws a random sample $\mathbf{z}$ from the distribution $\mathcal{N}(\mu,\sigma^2)$. This stochastic step introduces the non-determinism necessary to account for the multiple initial configurations that could have produced the same final momenta.
Next, the decoder takes the sample $\mathbf{z}$ and reconstructs a candidate molecular structure $\hat{\mathbf{x}}$. By sampling different $\mathbf{z}$ values from the distribution defined by the encoder, the model can generate the entire collection of valid initial configurations.

To learn the parameters of the VAE model, the encoder-decoder network is trained jointly using a dual-objective loss function, which is the sum of two terms: The first term is the MSE as defined in Eq.~\eqref{loss}, i.e., the same as used for the MLP, to ensure reconstruction accuracy, while the second term is the Kullback–Leibler divergence, which serves as a regularizer, encouraging the learned latent distribution to remain close to a standard normal distribution. This regularization organizes the latent space into a continuous, searchable manifold, allowing the network to learn the underlying statistics of the training data.

\subsection{Metrics}
\label{sec:metrics}

\textbf{Mean Squared Error (MSE) and Normalized Root Mean Squared Error (NRMSE).} The MSE is the average squared Euclidean norm between the true and predicted positions as defined by Eq.~\eqref{loss-all}. NRMSE is the normalized square root of the MSE, which allows for a direct comparison of model performance across datasets or variables with different numerical scales. Let $d$ denote the output dimension and $m$ the number of evaluated samples.
The scale-independent normalized RMSE error is defined as:

\begin{equation} \mathrm{NRMSE} = \frac{ \sqrt{MSE} }{ R  } \end{equation}
\noindent
where 
$\displaystyle R = \max_{i,d} x^{(i)}_d - \min_{i,d} x^{(i)}_d$.
Lower MSE and NRMSE values indicate improved predictive accuracy, with $\mathrm{MSE}=0$ or $\mathrm{NRMSE}=0$ corresponding to a perfect prediction.

\textbf{Coefficient of Determination (\(R^2\)).} Measures how well predicted positions explain the variance of the true positions:
\begin{equation}
    R^2 = \frac{1}{N_\text{atoms}}\sum_{a=1}^{N_\text{atoms}}\left[
    1 - \frac{\sum\limits_{i=1}^m \left\|\hat{\mathbf{r}}_{a,0}^{(i)} - \mathbf{r}_{a,0}^{(i)}\right\|_2^2}{\sum\limits_{i=1}^m \left\|{\mathbf{r}}_{a,0}^{(i)} - \overline{\mathbf{r}}_{a,0}\right\|_2^2}
    \right]
\end{equation}
where \(\bar{\mathbf{r}}_{a,0}\) is the mean of the true initial position of atom $a$ computed from the dataset being evaluated, and $\hat{\mathbf{x}}=(\hat{\mathbf{r}}_{1,0},\ldots,\hat{\mathbf{r}}_{N_{\text{atoms}},0})$ is the vector of predicted positions, as defined earlier in Section \ref{MLP}. 
$R^2=1$ corresponds to a perfect prediction, whereas a model that always predicts the mean position \(\bar{\mathbf{r}}_{a,0}\) yields $R^2=0$. A negative value for \(R^2\) can occur if the model predicts positions that are outside of the variance of the true positions.

\textbf{Average distance error} between predicted and true atom positions:
\begin{equation}
\mathrm{ADE} \;=\;  \frac{1}{N_{\mathrm{atoms}}}\sum_{a=1}^{N_{\mathrm{atoms}}}\big\|\hat{\mathbf{r}}_{a,0} - \mathbf{r}_{a,0}\big\|_2\,, 
\end{equation}
in units of atomic units (a.u.). 

\subsection{Evaluation Strategy and Computational Details} 
\label{sec:compdetails}
All models are trained using systematic hyperparameter optimization with the Optuna Python library. Model architectures and training parameters are selected based on validation-set performance, with the best configuration evaluated on the test set. The dataset was randomly partitioned into training, validation, and test subsets using a $70\%/15\%/15\%$ split, respectively, with samples assigned at random to each subset to ensure representative data distributions and to support unbiased model training, hyperparameter tuning, and final performance evaluation.

All experiments were conducted on a workstation equipped with an NVIDIA A40 GPU (46{,}068~MB VRAM) and an Intel(R) Xeon(R) Gold~6326 CPU @~2.90~GHz. The software stack consisted of PyTorch~2.7.1 with CUDA~12.6. We evaluated MLP and VAE models on 24-dimensional tabular datasets with sizes ranging from $2\times10^{4}$ to $1\times10^{7}$ samples. Optuna-based hyperparameter optimization was performed with up to 100 trials per experiment, exploring a wide range of network depths and hidden-layer sizes. When required, an additional identical A40 GPU was used to accommodate larger models.

The wall-clock training time for a single trial scaled with both dataset size and model capacity. For small datasets ($2\times10^{4}$ to $1\times10^{5}$ samples), training typically required $1$--$3$ minutes for MLP models and $3$--$6$ minutes for VAE models. For medium-scale datasets ($\sim10^{6}$ samples), training time increased to approximately $8$--$15$ minutes for MLPs and $12$--$25$ minutes for VAEs. For the largest datasets ($\sim10^{7}$ samples), a single trial required roughly $40$--$70$ minutes for MLPs and $60$--$100$ minutes for VAEs. Consequently, Optuna-based hyper-parameter optimization with up to 100 trials resulted in total runtimes ranging from several hours to multiple days, depending on dataset size and architectural complexity.

\section{Results and Discussion}
\label{sec:results}

\subsection{Model Benchmarking}
\label{benchmarking}

As the first step of our study, we designed a set of numerical benchmarking experiments to systematically evaluate our models and understand the effects of training data and architectural choices. Since the ultimate goal of the study is to apply the models to reconstruct the geometry of various isomers of the 5-atom bromochlorofluoromethane (CHBrClF) molecule (see Fig.\ \ref{fig:schematic}a), we start by generating datasets of arbitrary sample molecules comprised of 5 atoms (carbon, C; hydrogen, H; bromine, Br; chlorine, Cl; and fluorine, F), where each atom is randomly positioned in a sphere with a radius R=10\, atomic units (a.u.) or R=4\, a.u. To avoid cases where two atoms are unrealistically close to each other, which would lead to extremely large Coulomb repulsion, the minimum distance between two randomly placed atoms is set to be larger than 0.5\, a.u. We present a comparison using datasets of 1,000,000 and 10,000,000 samples generated from a uniform random distribution. As described in Section \ref{sec:compdetails}, all experiments employ a 70/15/15 split for training, validation, and testing, respectively. The validation set was used for hyperparameter tuning, while the model performance was evaluated on the heldout test set, which was not exposed during training.

The results are shown in Table \ref{tab:family_split}. Both the MLP and VAE architectures demonstrate comparable predictive performance for both dataset configurations. While the VAE consistently achieves a slightly lower MSE than the MLP, both exhibit similar \(R^2\) values. Note that the higher MSE values observed for the 10 a.u.\ sphere can be attributed to the larger spatial domain rather than the reduced training data density, as evidenced by the NRMSE values, which are the same for both datasets. Since the NRMSE expresses the prediction error relative to the overall dynamic range of a dataset, it normalizes out any scale-dependent effects, allowing for direct comparison between the 4~au and 10~au cases.
\begin{table}[t]
\centering
\caption{Performance comparison of a \textit{Multilayer Perceptron (MLP)} and a \textit{Variational Autoencoder (VAE)} evaluated on random-positions-in-sphere molecular datasets.}
\label{tab:family_split}
\begin{tabular}{l l r r r r}
\toprule[1.5pt]
\textbf{Dataset} & \textbf{Model} & \textbf{MSE} & \(\mathbf{R^2}\) & \textbf{ADE} & \textbf{NRMSE} \\
\midrule
4 au (10M)  & MLP & 0.33 & 0.94 & 1.48 & 0.04 \\
10 au (1M) & MLP & 2.69 & 0.92 & 4.53 & 0.04 \\
4 au (10M)  & VAE & 0.32 & 0.94 & 1.45 & 0.04 \\
10 au (1M) & VAE & 2.58 & 0.92 & 4.29 & 0.04 \\
\bottomrule
\end{tabular}
\end{table}

To validate robustness with respect to random initialization, we conducted experiments using 10 different random seeds; the resulting standard deviations of both MSE and $R^2$ for the two models were on the order of $10^{-5}$, indicating highly stable and reproducible performance across random seeds. Given this negligible variability, the effect of random initialization on model performance is effectively insignificant, and therefore, for the remainder of the experiments, we report results obtained using a single fixed random seed without loss of generality or statistical reliability. 

In order to test the reconstruction approach on an actual molecular structure, we simulate Coulomb explosion data for the equilibrium geometry of CHBrClF, with the initial positions sampled from a Gaussian distribution with $\sigma$ = 0.4 a.u.\ around the equilibrium geometry to approximately account for both the thermal spread of initial geometries as well as the additional broadening during the explosion process \cite{venkatachalam2025exploiting}. As before, the dataset is randomly split into training, validation, and test sets using a 70/15/15 ratio.

\begin{table}[b]
\centering
\caption{Evaluation of different models on real-world geometry initial positions sampled from a Gaussian distribution. 
}
\label{tab:structured_subsets}
\begin{tabular}{l r r r}
\toprule[1.5pt]
\textbf{Model} & \textbf{MSE} & \(\mathbf{R^2}\) & \textbf{ADE} \\
\midrule
Normal MLP & 0.01 & 0.98 & 0.25 \\
Normal VAE & 0.01 & 0.99 & 0.16 \\
\bottomrule
\end{tabular}
\end{table}
The results are shown in Table \ref{tab:structured_subsets} and display very high-accuracy reconstruction results. Both architectures achieve excellent performance on this realistic molecular geometry dataset, with \(R^2\) values exceeding 0.98. The VAE demonstrates superior MSE performance with a value of 5.96$\times 10^{-3}$, while the MLP achieves a slightly higher \(R^2\) value than the MLP and an MSE of 9.25$\times 10^{-3}$.

\subsection{Application to Coulomb Explosion of a Mixed Sample of CHBrClF Isomers}
\label{CHBrClF}
\begin{figure*}
    \centering
    \includegraphics[width=2\columnwidth]{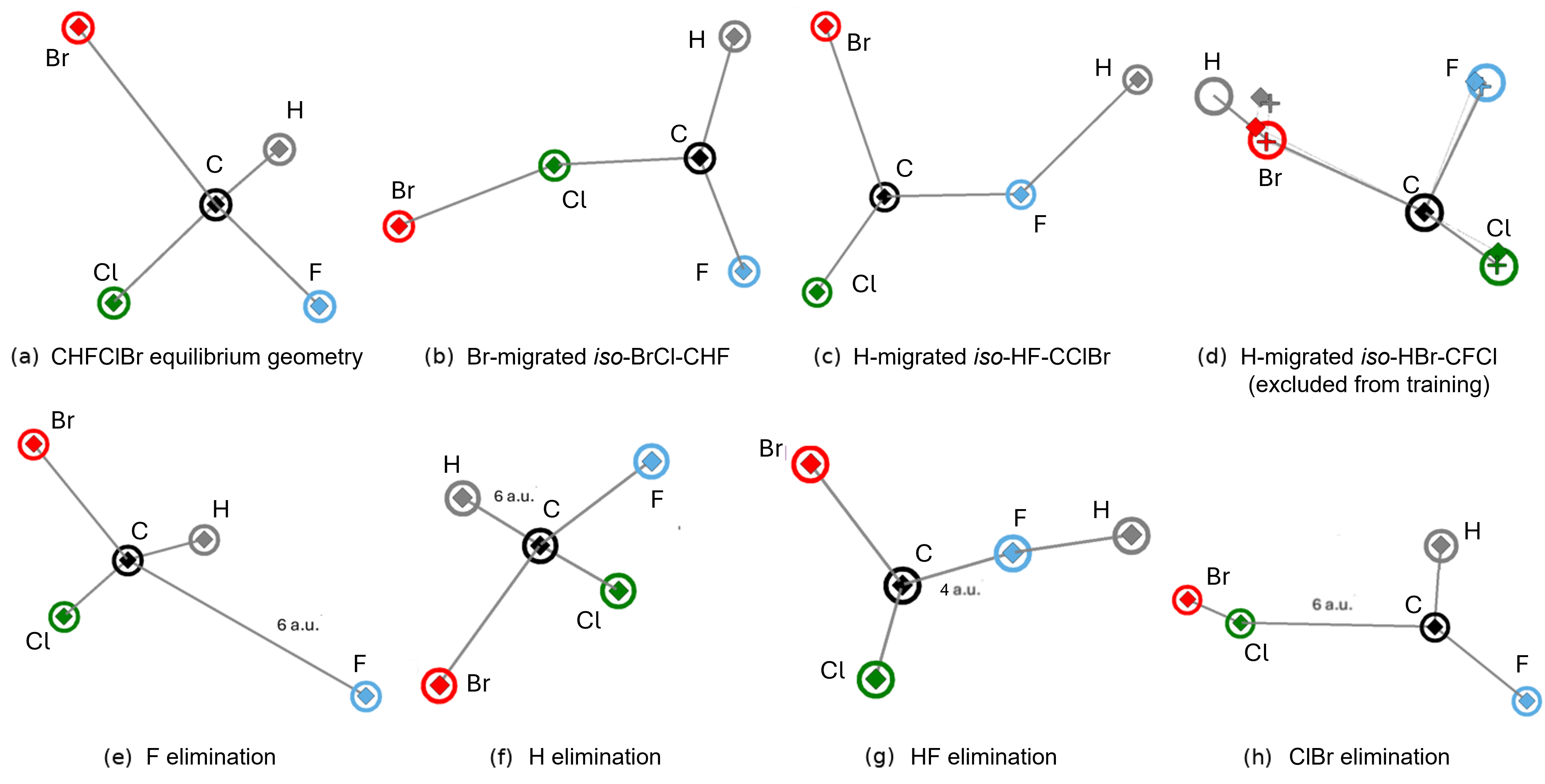}
    \caption{Ball-and-stick-models of the eight CHBrClF isomers used to test the real-space reconstruction model. The open circles mark the center of mass of the initial position-space distribution used as a starting point for the Coulomb explosion simulations (''ground truth"). The diamonds mark the center of mass of the retrieved position-space distribution. For the results shown here, isomer (d) was not part of the training dataset. The crosses in (d) mark the results for the case where the random-positions-in-sphere data was added to the training data.}
    \label{fig:results}
\end{figure*}
In order to test how the reconstruction performs when several isomers are present, e.g., as competing product channels in a pump-probe experiment, we include eight distinct geometric isomers in both training and test sets to evaluate the ability of the models to distinguish subtle structural differences. The isomers, shown as ball-and-stick structures in Fig.\ \ref{fig:results}, represent idealized reaction products of a fictitious pump-probe experiments, including (a) the original CHBrClF equilibrium geometry, (b) a Br-migrated \textit{iso}-BrCl-CHF form, (c) H-migrated \textit{iso}-HF-CClBr, (d) H-migrated \textit{iso}-HBr-CFCl, (e) a dissociation into CHClBr + F (F elimination), (f) a dissociation into CFClBr + H (H elimination), (g) a two-body breakup into CHF + ClBr (ClBr elimination), and (h) a two-body breakup into CClBr + HF (HF elimination). Note that CHBrClF in its equilibrium geometry is chiral. For simplicity, we chose only one of the two enantiomers for our study and focused on other structural isomers, but it was shown previously that CEI can also distinguish chiral enantiomers \cite{pitzer2013direct}.

While not all of the structures listed above may indeed be formed in a realistic pump-probe experiment on CHBrClF, they represent a reasonable sample of the structural motifs one would expect as possible reaction products after photoexcitation of polyhalomethanes. Each of the structures shown in Fig.\ \ref{fig:results} was first broadened by a Gaussian distribution with standard deviation $\sigma$ = 0.4 a.u.\ to create the training and validation data. A slightly ''tighter" distribution with 0.5 $\sigma$ was then used to create the test data to avoid sampling the edges of the training space.

Initially, the model was trained on a dataset containing all eight isomers, and the reconstruction of each isomer's initial positions was evaluated, yielding extremely low MSE and high R$^2$ values and accurate reconstruction results (see Table \ref{tab:isomers_combined_summary}). These results demonstrate that the model is capable of learning structural features that allow it to not only recover molecular geometries with high fidelity but also reliably distinguish between the different isomers within the dataset.
\begin{table}[b]
\centering
\caption{
Reconstruction performance of the \textit{Multilayer Perceptron (MLP)} and \textit{Variational Autoencoder (VAE)} models for three experimental settings: 
(1) all-isomer training (8$\rightarrow$8) with all 8 isomers included in the test set, 
(2) leave-one-isomer-out training (train on Isomers (a)--(c), (e)--(h); test on the left-out Isomer (d)) without augmentation, and 
(3) leave-one-isomer-out training with data augmentation (test on the left-out Isomer (d)).
}
\label{tab:isomers_combined_summary}
\begin{tabular}{l r r r}
\toprule[1.5pt]
\textbf{Model / Experiment} & \textbf{MSE} & \(\mathbf{R^2}\) & \textbf{ADE} \\
\midrule
All-isomer (8$\rightarrow$8) -- MLP & 0.01 & 0.99 & 0.22 \\
All-isomer (8$\rightarrow$8) -- VAE & 0.01 & 0.99 & 0.09 \\
\hline
Leave-one-out (Isomer d), no aug. -- MLP & 6.44 & -0.25 & 8.50 \\
Leave-one-out (Isomer d), no aug. -- VAE & 0.41 & -0.02 & 0.96 \\
\hline
Leave-one-out (Isomer d), aug. -- MLP & 0.42 & 0.09 & 2.05 \\
Leave-one-out (Isomer d), aug. -- VAE & 0.32 & 0.12 & 0.76 \\
\bottomrule
\end{tabular}
\end{table}

\begin{figure}[h!]
    \centering
    \includegraphics[width=0.9\linewidth]{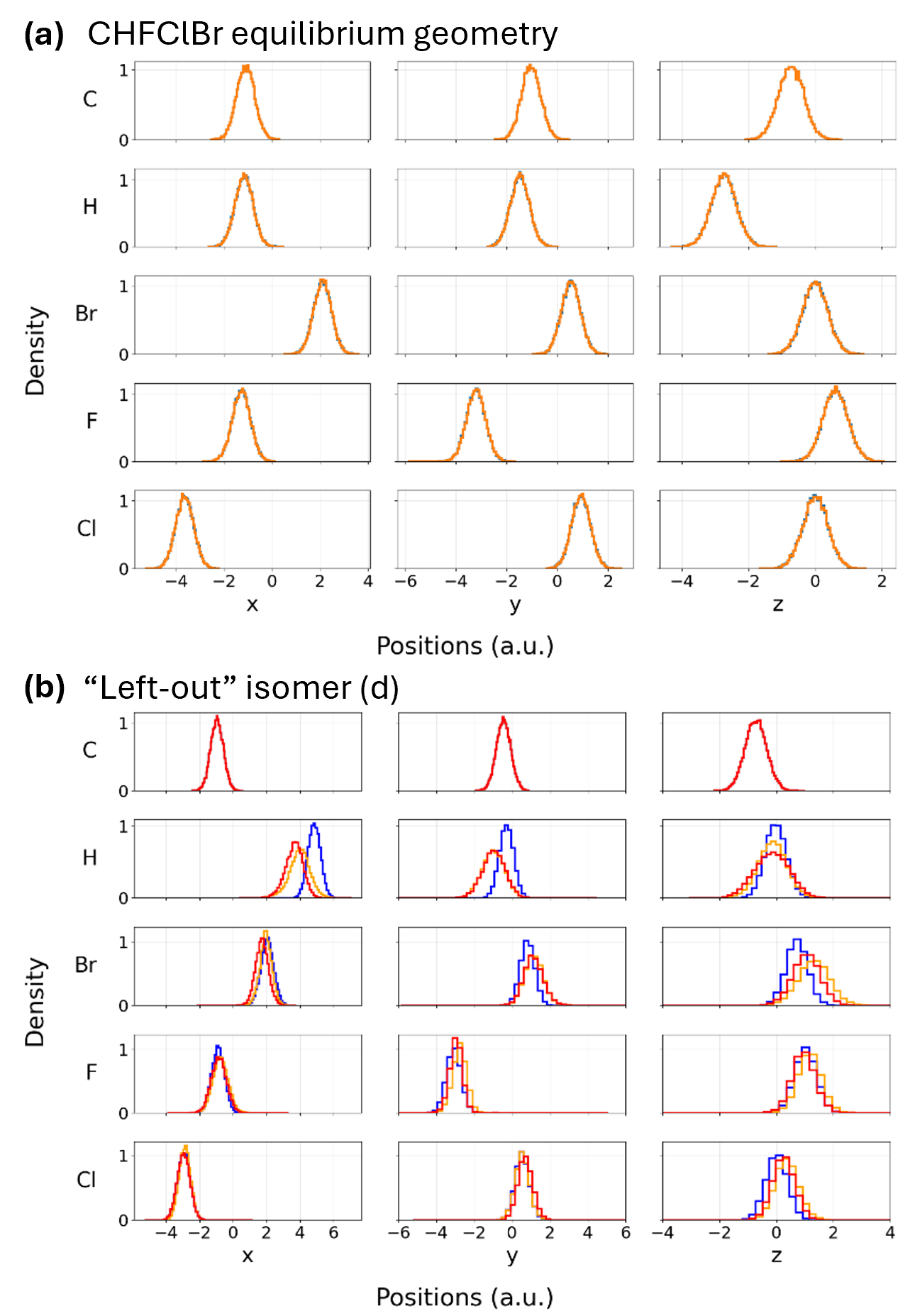}
    \caption{Initial (blue) and predicted (orange) position distributions for all atoms of (a) the original CHBrClF molecule and (b) for the isomer that was left out of the training data. In panel (b), the red lines show the results when the training data was augmented by the random-positions-in-sphere data. Similar plots for all other isomers are shown in the Supplementary Online Information.  
    }
    \label{fig:distr}
\end{figure}

Subsequently, we left out isomer (d) from the training data in order to model a scenario where an unexpected reaction product is formed that was not foreseen when building the training data and assess reconstruction on this unseen case. The results from this scenario are shown as solid circles in Fig.\ \ref{fig:results}. Isomers (a) through (c) and (e) through (h) are still reconstructed with high accuracy, whereas the ''unexpected" isomer (d), that was not included in the training data, is still reconstructed as a distinctly different structure, albeit with a significant deviation from the ground truth, in particular with respect to the position of the hydrogen atom, which is most different in this structure as compared to the other isomers that were included in the training data. In this setting, the reconstruction yielded $R^2$ values of $-0.25$ for the MLP and $-0.02$ for the VAE, as shown in Table~\ref{tab:isomers_combined_summary}. Interestingly, this deviation mostly shows up in only one coordinate of the H positions (see Fig.\ \ref{fig:distr}). It is important to note, however, that the reconstruction of the ''unexpected" isomer (d) results in a geometry that is close enough to the ground truth that the nature of this unexpected product as being an isomer where the H has migrated towards the Br ligand can be guessed from the reconstruction. If this was the case in the reconstruction of real experimental data with unknown constituents, the Coulomb explosion pattern for this isomer could thus be simulated and included in an improved training dataset for a second iteration of the reconstruction.

Finally, in order to investigate if the reconstruction of unexpected products can be improved, we used the 7-isomer training dataset augmented by the random-positions-in-sphere data from our earlier tests, in which all five atoms were placed at arbitrary positions, in order to assist with the generalization to unseen isomers. This augmentation introduces greater structural diversity into the training set and significantly improves the reconstruction of isomer (d) using the MLP and slightly improves the reconstruction using the VAE, as shown by the improved MSE and $R^2$ values listed in Table~\ref{tab:isomers_combined_summary} and by the red lines in Fig.\ \ref{fig:distr}). Interestingly, it can be seen from Fig.\ \ref{fig:results}d that the improvement occurs primarily for the F, Cl, and Br atoms and less for the H atom, whose reconstructed position differs the most from the ground truth.

\begin{table}[b]
\caption{Average distance errors (ADE) for each atom for different training settings for the VAE.}
\label{tab:atomic_distances}
\begin{tabular}{l r r r r}
\toprule
\textbf{Training Setting} & \textbf{Br} & \textbf{Cl} & \textbf{F} & \textbf{H} \\
\midrule
All-isomer                     & 0.11 & 0.08 & 0.09 & 0.07 \\
Leave-one-isomer-out           & 0.77 & 0.61 & 0.84 & 1.62 \\
Leave-one-isomer-out + aug.  & 0.69 & 0.43 & 0.44 & 1.50 \\
\bottomrule
\end{tabular}
\end{table}
To further evaluate the reconstruction quality, we analyzed atomic distance errors for each atom across all three training settings, as shown in Table \ref{tab:atomic_distances}. When all eight CHBrClF isomers were included in training, the ADE errors were around 0.1 a.u.\ for all atoms, indicating robust recovery of molecular geometries. In the leave-one-isomer-out case, where isomer (d) was excluded during training and used only for testing, errors increased substantially, with the hydrogen atom showing the largest deviation. This reflects the fact that the hydrogen position in isomer (d) was not represented in the training data. Finally, when random molecular configurations were added to the training set alongside the seven isomers, all position errors including the hydrogen position error decreased, suggesting that the additional structural diversity partially improved reconstruction. Nevertheless, the hydrogen position remained the most challenging one to reconstruct accurately, highlighting the difficulty of capturing novel atomic arrangements not encountered during training.  

\section{Conclusion}
We have shown that accurate molecular structure retrieval from simulated Coulomb explosion imaging data is possible by employing a neural network-based approach to solve the underlying inverse problem and infer the initial atomic positions from the final fragment-ion momenta. Although the average per-atom position error of approximately 0.1 atomic units (0.05 \AA), corresponding to less than 5\% of the typical bond lengths in the molecules we studied, is significantly larger than what high-resolution structure determination techniques such as X-ray or electron diffraction typically achieve for the equilibrium structure of a molecule, it is sufficient for unambiguous distinction and identification of the eight different molecular species in our mixed test sample.

We have compared a Multilayer Perceptron (MLP) and a Variational Autoencoder (VAE) across multiple datasets and found that both architectures achieve high reconstruction accuracy when all structures were included in the training data,
with the VAE demonstrating superior generalization capability, particularly when reconstructing unseen molecular isomers, as evidenced by its significantly better performance in the leave-one-isomer-out experiments. 
Overall, we conclude that although detailed performance varies by metric and dataset, the differences in reconstruction accuracy between models are less significant than the impact of selecting appropriate training data, with data augmentation using random molecular configurations proving particularly effective for improving generalization to unexpected structures.

By applying the models to simulated Coulomb explosion imaging data for several polyhalomethane isomers, we have shown that accurate structure retrieval is possible even when many competing molecular structures are present in the data, which would often be the case in a realistic pump-probe experiment, and which can pose significant challenges to ensemble-averaging techniques such as X-ray and electron diffraction. Furthermore, we have shown that a qualitatively meaningful structure is retrieved even if the corresponding isomer is not included in the training data. This makes it possible to identify unexpected reaction products and to iteratively improve the reconstruction accuracy by including the corresponding structure in the training data in the next iteration.

A next step in this endeavor would aim to enable molecular structure retrieval based on experimental data rather than simulated data. In principle, if the molecular explosion process can be more precisely simulated, structure retrieval could be formulated as an optimization problem, training the model on simulated data and then extended to experimental data \cite{Li2025arxiv,guo2025decoding}. Presently, the primary limitation for accurate simulations is the insufficient understanding of the ionization and charge buildup process. Additionally, deviations of potential energy surfaces from purely Coulombic behavior at small internuclear distances, even for highly charged ions, further complicate accurate modeling.

For the case of X-ray multiphoton ionization with an X-ray free-electron laser, the charge buildup process involves a complex interplay of sequential inner-shell ionization, ultrafast decay, intramolecular charge redistribution, and nuclear dynamics~\cite{Rudenko2017,boll2022x,li2021simple,li2022coulomb}, which can, in principle, be modeled with reasonably high accuracy. However, the computational expense of these calculations is currently very high \cite{boll2022x,Li2025arxiv}. For the case of strong-field ionization with intense near-infrared laser pulses, accurate modeling of the multiple ionization dynamics and the concurrent nuclear dynamics remains very challenging \cite{Ashrafi-Belgabad2024}, and we are not aware of any such work on molecules of size and complexity similar to the 5-atom halogenated methane that was used as a showcase example in the present study.

Since new high-average-power, high-repetition-rate laser sources are now making it possible to record multi-coincidence CEI data for a given, static molecule in a matter of minutes rather than many hours, an alternative approach could be to build a sufficiently large training data base of experimental CEI data for known molecular structures to be able to incorporate this data in the training of the neural network.

Finally, we would like to point out that since our CEI reconstruction is performed on a molecule-by-molecule basis, it has the potential to recover not only average atomic positions but also the square of the nuclear wavefunction, i.e., the probability density distribution, making CEI combined with accurate molecule-by-molecule geometry retrieval and powerful technique for molecular wavefunction imaging. While reconstruction of \textit{average} molecular geometries is also possible with ''incomplete" coincidence or covariance data acquired and averaged over many laser shots\cite{Li2025arxiv,guo2025decoding}, wavefunction imaging requires kinematically complete fragment ion coincidence measurements, where \textit{all} fragment ions are detected in coincidence on a shot-by-shot basis. Such complete coincidence detection has recently been demonstrated for molecules consisting of up to eight atoms\cite{venkatachalam2025exploiting}, and the increasing availability of high-repetition-rate laser, combined with the development of new detectors with higher multi-particle detection efficiencies, will almost certainly push this limit to even larger molecules in the near future.

\section*{Acknowledgments}
This work was supported by a GRIPex award from Kansas State University. We acknowledge helpful discussions about inverse problems with Prof.\ Dinh Liem Nguyen, about neural networks with Prof.\ William Hsu, and about the application of ML/AI to Coulomb explosion imaging with Dr.\ Xiang Li and Dr.\ Enliang Wang in the early stages of the project. H.V.S.L., A.R., L.G., and D.R. are funded through the Chemical Sciences, Geosciences, and Biosciences Division, Office of Basic Energy Sciences, Office of Science, U.S.~Department of Energy under grant no.~DE-FG02-86ER13491.  

\section*{Data Availability Statement}
All code and datasets supporting this research are publicly available. The implementations of the Multilayer Perceptron (MLP) and Variational Autoencoder (VAE) models are available on GitHub at \hyperlink{https://github.com/Amir31337/CEI}{https://github.com/Amir31337/CEI}. The repository includes training and evaluation scripts for all experimental configurations: random sphere distributions (1,000,000 samples at R=10 a.u.\ and 10,000,000 samples at R=4 a.u.), real-world molecular geometries sampled from Gaussian distributions, and all eight isomer configurations with leave-one-isomer-out experiments. Additionally, scripts are provided to generate the datasets used in all experiments, including the random distribution data, normal-broadened molecular geometries with $\sigma$ = 0.4 a.u., and the full set of isomer data with and without data augmentation.

\bibliographystyle{aipnum4-1}
\bibliography{references}
\appendix







\begin{figure*}[h!]
    \centering
    \includegraphics[width=0.85\linewidth]{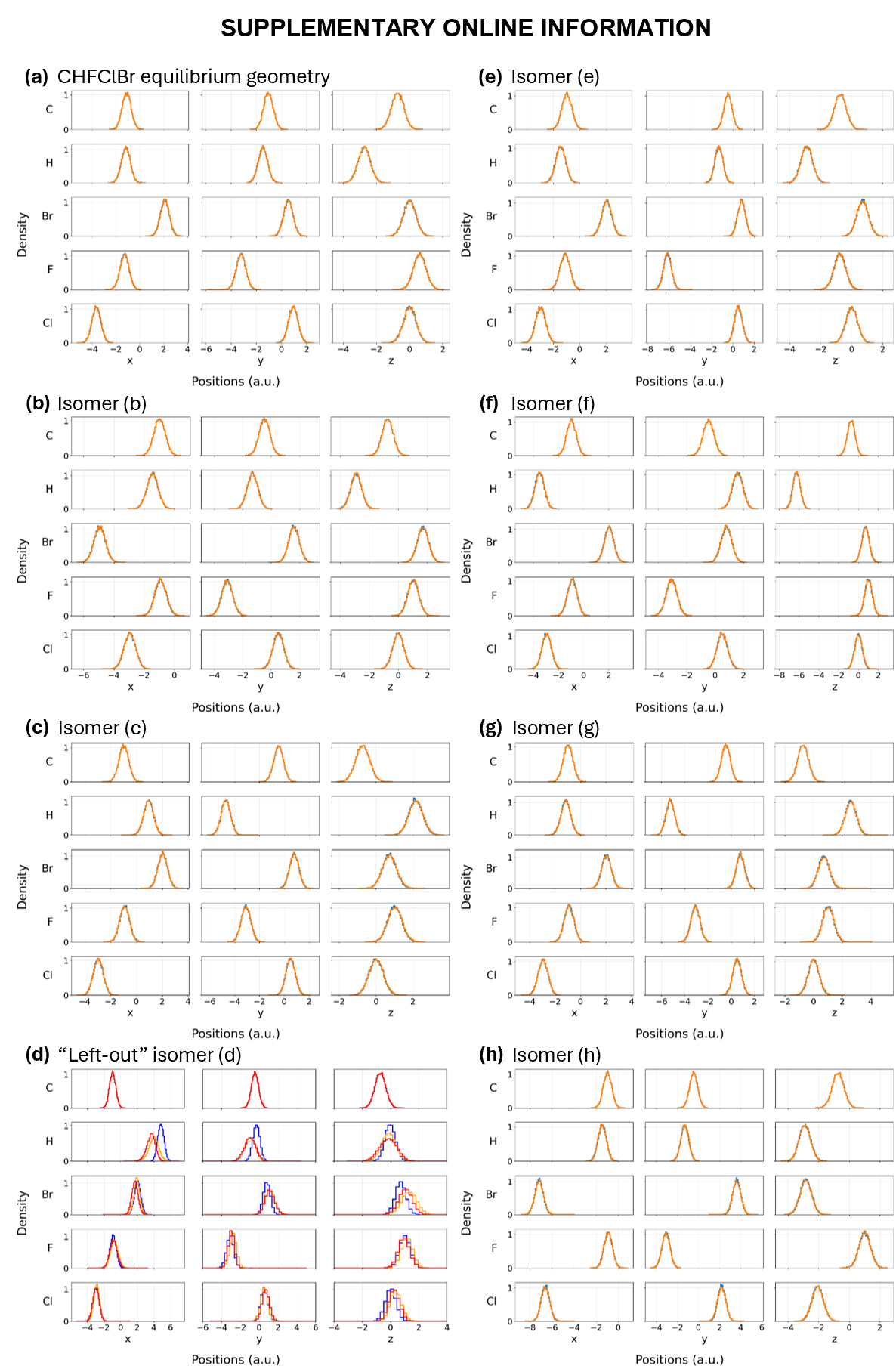}
    \caption{
    Initial (blue) and predicted (orange) position distributions for all atoms of the eight tested isomers, shown in the same order as in Fig.\ \ref{fig:results} of the main text. In panel (d), the red lines show the results when the training data was augmented by the random-positions-in-sphere data.}
    \label{fig:distr_all}
\end{figure*}

\end{document}